\documentstyle[prl,aps,twocolumn,epsfig]{revtex}

\begin{document}
\twocolumn[\hsize\textwidth\columnwidth\hsize\csname@twocolumnfalse\endcsname
\title{Dynamical Anomalies and Intermittency in Burgers Turbulence}
\author{Michael L\"assig}
\address{
Max-Planck-Institut f\"ur Kolloid- und Grenzfl\"achenforschung,
Kantstr.~55, 14513 Teltow, Germany\\
Institut f\"ur theoretische Physik, Universit\"at zu K\"oln,
Z\"ulpicher Str. 77, 50937 K\"oln, Germany}

\date{November 6, 1998}
\maketitle

\begin{abstract}
We analyze the field theory of fully developed Burgers turbulence. Its key
elements are {\em shock} fields, which characterize the singularity 
statistics of the velocity field. The shock fields enter an operator product 
expansion describing {\em intermittency}. The latter is found to be constrained 
by {\em dynamical anomalies} expressing finite dissipation in the inviscid
limit. The link between dynamical anomalies and intermittency is argued to 
be important in a wider context of turbulence. 
\vspace{10 pt}

PACS numbers: 64.60.Ht, 68.35.Fx, 47.25Cg
\vspace{24pt}
\end{abstract}

%\pacs{PACS numbers: 00.00.xx}

\vfill
] %matches %\twocolumn
\narrowtext %\newpage

A field theory of hydrodynamic turbulence is difficult in two ways: 
It is far from equilibrium
and far from the realm of standard perturbative 
renormalization~\cite{Frisch.book}. New non-perturbative concepts have
been established for simpler model systems
sharing some characteristics of turbulent fluids. In particular, 
the stochastic Burgers equation
\begin{equation}
\partial_t {\bf v} + ({\bf v} \cdot {\bf \nabla}) {\bf v}
          = \nu \, {\bf \nabla}^2 {\bf v}  + {\bf \nabla} \eta 
\label{Burgers}
\end{equation}
has become an important model for turbulence, recently studied 
by a variety of 
methods~\cite{Sinai,ChekhlovYakhot,BouchaudAl,Polyakov,GurarieMigdal96,%
BalkovskyAl,Boldyrev,EAl,GotohKraichnan,Bec}.
It governs the time evolution of a vortex-free velocity field
${\bf v} ({\bf r},t) = \nabla h({\bf r},t)$. 
The equivalent scalar equation 
\begin{equation}
\partial_t h = \nu \nabla^2 h + \frac{1}{2} (\nabla h)^2 + \eta
\label{KPZ}
\end{equation}
is known as the Kardar-Parisi-Zhang equation~\cite{kpz,Lassig.review}. 
The driving potential $\eta({\bf r},t)$ is 
Gauss distributed with mean $\overline \eta ({\bf r},t) = 0$ and 
correlations 
\begin{equation}
\overline{\eta({\bf r},t) \eta({\bf r}',t')} 
     = \varepsilon r_0^2 \delta (t - t') G(|{\bf r} - {\bf r}'|/ r_0) 
\label{etaeta}        
\end{equation}
over a characteristic scale $r_0$. The function $G$ is taken to be
analytic with $G(\rho) = 1 - G_2 \rho^2 + O(\rho^4)$ and $G_2 > 0$. 
There is a second characteristic scale, the dissipation length 
$a_0 \equiv \nu^{3/4} \varepsilon^{-1/4}$. Burgers turbulence occurs
at high values of the Reynolds number $R \equiv (r_0/a_0)^{4/3}$ 
and shows strong {\em intermittency}. For 
example, the longitudinal velocity difference moments of order 
$k = 2,3,\dots$ take the form 
\begin{equation}
\langle [v_{\|} ({\bf r}_ 1) - v_{\|} ({\bf r}_ 2)]^k \rangle \sim
   |{\bf r}_1 - {\bf r}_2|^{-k x_{{\bf v}} + \tilde x_k} 
   \, r_0^{- \tilde x_k}
\label{vk}
\end{equation}
in the inertial scaling regime 
$r_0 / R \ll |{\bf r}_1  - {\bf r}_2| \ll r_0 
$~\cite{ChekhlovYakhot,BouchaudAl}. The third moment grows linearly
with $|{\bf r}_1  - {\bf r}_2|$, i.e., 
$x_{\bf v} = - 1/3$ and $\tilde x_3 = 0$. The other moments, however,
acquire a singular dependence on $r_0$, which defines the  
intermittency exponents $\tilde x_k$.  
The turbulent state is associated with a particular strong-coupling limit 
of Eq.~(\ref{Burgers}), called the turbulent limit in the sequel:
$\nu \to 0$  at fixed driving given by~(\ref{etaeta}). In that limit,
the velocity field acquires discontinuities called {\em shocks} (see Fig.~1).
It is these singularities that cause intermittency.

Burgers' equation has a number of further applications. It
has been proposed as a model for galaxy formation~\cite{Zeldovich}. 
The Kardar-Parisi-Zhang equation (\ref{KPZ}) models stochastic surface 
growth described by the ``height''
field $h({\bf r},t)$, and is related to directed polymers in a quenched random 
medium. In these contexts, the driving force is usually taken to be
random in time and space, 
\begin{equation} 
\overline{\eta({\bf r},t) \eta({\bf r}',t')} 
     = \gamma^2 \delta (t - t') \delta({\bf r} - {\bf r}') \;.
\label{etaeta2}
\end{equation}
For $\nu^3/ \gamma^2 \to 0$, there is again a strong coupling limit,
which, however, is quite different from the turbulent limit. The scaling is 
probably non-intermittent, 
\begin{equation}
\langle (h({\bf r}_1) - h({\bf r}_2))^k \rangle 
  \sim |{\bf r}_1 - {\bf r}_2|^{-k x_h} \;.
\label{hk}
\end{equation}
The exponent $x_h$ depends on the dimension $d$. 
The value $x_h = -1/2$ for $d = 1$ is well-known~\cite{FNS},
while $x_h = -2/5$ for $d = 2$ has been obtained only 
recently~\cite{Lassig.kpz98,Lassig.review}. 

This Letter studies the field theory of stationary Burgers turbulence.
The basic quantities of this field theory are scaling fields
containing powers of the local velocity ${\bf v}$ and its gradients, 
$\nabla {\bf v}$, $\nabla \nabla {\bf v}$, etc. In particular, 
the statistics of
the shocks is represented by a family of renormalized {\em shock fields}
$\Sigma_k$ ($k = 1,2, \dots$) that remain finite in the turbulent limit.  
Another family of fields $s_k$ describes velocity gradients away from
shocks. 
The short-distance properties of the scaling fields are encoded in an 
{\em operator product expansion} (OPE). This is a familiar concept in 
field theory (see, e.g., ref.~\cite{Cardy.book}). It has recently been
\begin{figure}[b]
\vspace*{-1.2cm}
\epsfig{file=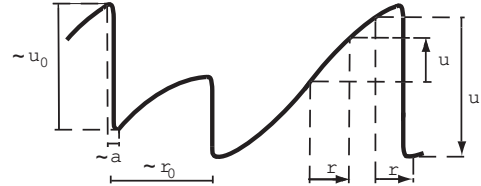, height=6.0cm, angle=0}
\vspace{-1.8cm}
\caption{\small 
Stationary Burgers turbulence in one dimension. The velocity profile
$v (r, t)$ at a given time $t$ is an ensemble of shocks with 
amplitude $\sim u_0$, width $\sim a$, and mutual distance $\sim r_0$.   
The probability distribution $P_r (u)$ of the longitudinal velocity 
difference over a distance  $a \ll r \ll r_0$
is generated by no-shock and  single-shock configurations
(lines with arrows).}
\end{figure}  
\noindent
extended to non-equilibrium systems. OPEs have been 
proposed for general turbulent systems by 
Adzhemyan, Antonov, and Vasil'ev~\cite{Petersburg}, 
Eyink~\cite{Eyink} (see also 
Duplantier and Ludwig~\cite{DuplantierLudwig}), 
and by L'vov and Procaccia~\cite{LvovProcaccia},
for Burgers turbulence by Polyakov~\cite{Polyakov},
and for the Kardar-Parisi-Zhang equation by L\"assig~\cite{Lassig.kpz98}.
The Burgers OPE discussed here differs from Polyakov's conjecture
by the explicit inclusion of the fields $\Sigma_k$ and $s_k$.
It is an important tool to understand the dynamics of Burgers turbulence.
The shock singularities 
preventing a straightforward evaluation of the equation of motion,
the stationary state is maintained by a rather subtle balance of driving 
forces, convection, and dissipation. We discuss the
resulting distribution of velocity differences (the functional form of which 
has been much  
debated~\cite{ChekhlovYakhot,BouchaudAl,Polyakov,GurarieMigdal96,%
BalkovskyAl,Boldyrev,EAl,GotohKraichnan,Bec}), as well as
the flux of the energy density and its generalizations.  
The energy dissipated per unit of volume and time remains 
finite in the turbulent limit $\nu \to 0$. In the 
field theory of Burgers' equation, this is reflected by 
{\em dynamical anomalies}, i.e., asymptotically finite dissipation terms 
in effective equations of motion for the inertial regime.
These are associated with 
conservation laws that are valid for the inviscid equation without 
driving ($\nu = 0$, $\varepsilon = 0$) but are broken in the driven state
at any finite $\nu$~\cite{Gurarie95,Polyakov}. 
We find anomalies given by
operator products involving the shock fields. This is not surprising since
dissipation takes place at the shocks. Dynamical anomalies and intermittency 
are indeed closely related: they are both generated by the singularities
of the velocity field. This conceptual link and the underlying theoretical
framework are expected to extend to other turbulent systems,
as we briefly discuss at the end of this Letter. 

The basic phenomenology of stationary Burgers turbulence is well established 
in one dimension~\cite{ChekhlovYakhot,GotohKraichnan} 
but appears to be much the same in higher 
dimensions~\cite{BouchaudAl,GurarieMigdal96}. 
The velocity profile ${\bf v}({\bf r}, t)$ at a given time $t$ looks similar to that
of decaying Burgers turbulence with random initial 
data~\cite{Kida,SheAl,GurbatovAl}.
It consists of {\em ramps} (i.e., regions where the velocity derivatives 
$\nabla {\bf v}$, $\nabla \nabla {\bf v}$, etc.~are finite)
separated by shock singularities, as 
sketched in Fig.~1. Shocks develop out of ramp regions through 
{\em preshock} singularities (which have the cubic root form 
$v({\bf r}) - v({\bf r}_{\rm ps}) \sim 
 - (\epsilon ({\bf r} - {\bf r}_{\rm ps}))^{1/3}$ 
for $d = 1$~\cite{EAl}). 
The shocks have  amplitudes of order
$u_0 \equiv (\varepsilon r_0)^{1/3}$ and distances of 
order $r_0$ to their neighbors;
the slopes of the ramps are of order $u_0/ r_0$. These 
scales are independent of $\nu$, while the typical shock width
$a \equiv r_0 / R = \nu/u_0$ vanishes in the turbulent limit. From these
shock characteristics, the velocity statistics can be inferred in 
an approximate way that appears to give the correct scaling.
Consider, for example, the longitudinal velocity difference 
$u \equiv v_\| ({\bf r}_1,t) - v_\| ({\bf r}_2,t)$
(with $v_\| \equiv {\bf v} \cdot ({\bf r}_1 - {\bf r}_2)/r$ and 
$r \equiv |{\bf r}_1 - {\bf r}_2|$) and the local excess velocity 
${\bf w} ({\bf r},t) \equiv {\bf v} ({\bf r},t) - \langle {\bf v} \rangle$.
These have normalized probability distributions 
 $P_r (u) {\rm d} u$ and $Q({\bf w}) {\rm d}^d {\bf w}$, respectively. 
$Q({\bf w})$ depends only on the absolute value $|{\bf w}|$ 
by rotational invariance
and parity but $P_r(u)$ is an asymmetric function of $u$ since the dynamics
is not invariant under the transformation ${\bf v} \to -{\bf v}$. 
Both distributions are invariant under Galilei transformations 
${\bf v}({\bf r},t) \to {\bf v}({\bf r} - {\bf v}_0 t, t) - {\bf v}_0$. 
In the turbulent limit, they are independent of $\nu$, i.e., of the shock 
width $a$. Hence, they can be written in scaling form,
\begin{equation}
P_r (u) = \frac{1}{u_r} \,
         {\cal P} \!\left ( \frac{u}{u_r}, 
                            \frac{r}{r_0} \right ) , \;\;\;
Q({\bf w}) = \frac{1}{u_0^d} \,
         {\cal Q} \!\left ( \frac{|{\bf w}|}{u_0} \right ) ,
\label{calP} 
\end{equation}   
where $u_r \sim u_0 r / r_0$ is the average velocity increment over
a distance $r$ on a ramp. By Eq.~(\ref{calP}), the 
single-point moments scale in a simple way,
\begin{equation}
\langle |{\bf w}|^k \rangle \sim u_0^k \sim r_0^{-x_k} \;,
\;\;\; 
x_k = - k/3 \;.
\label{dvk}
\end{equation}
For $\rho \equiv r/r_0 \ll 1$, the 
powers of the velocity difference 
can be expanded in the number 
of shocks present between ${\bf r}_1$ and ${\bf r}_2$. One obtains 
$u^k \sim p_0 u_r^k + p_1 u_0^k + O(p_2)$
with the $n$-shock probabilities $p_0 \simeq 1 - \rho$,
%with the zero- and single-shock probabilities  
$p_1 \simeq \rho + O(\rho^2)$, $p_n = O(\rho^n)$ .
This implies the {\em bifractal} 
moments~\cite{ChekhlovYakhot,BouchaudAl}
\begin{equation}
\langle u^k \rangle \sim \left \{
\begin{array}{ll}
u_r^k \sim r^k r_0^{-2k/3}              & \;\; (k < 1)
\\
(r / r_0) u_0^k  \sim r r_0^{(k - 3)/3} & \;\; (k > 1)
\end{array} \;. \right.
\label{uk}
\end{equation}

What does all this mean for the field theory of the turbulent state?
Correlation functions in the inertial regime should be represented 
by a field theory with short-distance cutoff $a$. Scale invariance 
emerges in the turbulent continuum limit $a \to 0$. Since that limit
is nonsingular for velocity correlations, the scaling dimension of
the field ${\bf v}({\bf r},t)$ takes the Kolmogorov value $x_{\bf v} = -1/3$;
the fields ${\bf v}^k ({\bf r},t)$ have dimensions $x_k = k x_{\bf v}$
($k = 1,2, \dots)$.  
Indeed, the distributions (\ref{calP}) and the resulting moments
(\ref{dvk}), (\ref{uk}) are covariant under the scale transformations
${\bf r} \to b {\bf r}$, 
$r_0 \to b r_0$, 
${\bf v} \to b^{x_{\bf v}} {\bf v}$
at fixed $\varepsilon$.
Turbulence does generate short-distance singularities for the 
moments of velocity {\em gradients}. Defining the longitudinal gradient 
$\nabla \cdot {\bf v} \equiv u/r$ with a discretization length $r \sim a$,
we write 
\begin{equation}
(\nabla \cdot {\bf v})^k ({\bf r},t) = 
    Z_k^{-1} \Sigma_k ({\bf r},t) + s_k ({\bf r},t) \;.
\label{singreg}
\end{equation}
The fields $\Sigma_k ({\bf r},t)$ and $s_k({\bf r},t)$ represent the 
contributions from configurations with and without a shock in the discretization
interval. Using Eq.~(\ref{uk}), we have
\begin{eqnarray}
Z_k^{-1} \langle \Sigma_k \rangle & \sim &
\frac{u_0^k}{ a^{k - 1}  r_0} \sim \frac{r_0^{- \tilde x_k}}{a^{k - 1}} \;,
  \;\;\; 
  \tilde x_k = - \frac{k - 3}{3} \;,
\label{sigmak}
\\
\langle s_k \rangle_{\rm reg} & \sim &(u_0/r_0)^k \sim r_0^{-x'_k} \;,
  \;\;\;
  x'_k = 2k/3 \;.
\label{sk}  
\end{eqnarray}  
The multiplicative renormalization factors 
$Z_k \sim a^{k - 1}$ absorb the short-distance singularities in (\ref{sigmak}),
which defines the cutoff-independent fields
$\Sigma_k$ of dimension $\tilde x_k$. The slope fields $s_k$ 
have different dimensions $x_k'$ determined by the 
regular part (\ref{sk}) of the ramp slope moments. 

We now make the assumption that the velocity field satisfies an 
OPE of the form
\begin{eqnarray}
{\bf v}({\bf r}_ 1) \dots {\bf v}({\bf r}_ k) & = &  
     r^{- k x_{\bf v} + \tilde x_k} \, C_k \,
     \Sigma_k ({\bf r}_1) + 
     \nonumber \\
     & &  r^{- k x_{\bf v} + x_k'} \, c_k \,
     s_k ({\bf r}_1) + \dots 
\label{vope}
\end{eqnarray}
with $r \equiv |{\bf r}_1 - {\bf r}_2|$. 
Both sides of this relation are tensors of rank $k$ whose indices are 
suppressed. The right hand side is a sum over all local scaling fields.
The displayed terms contain the lowest-dimensional Galilei-invariant 
fields representing shock and ramp configurations, respectively, 
multiplied by dimensionless coefficient functions 
$C_k ({\bf r}_{12}/r, \dots, {\bf r}_{1k}/r )$ and
$c_k ({\bf r}_{12}/r, \dots, {\bf r}_{1k}/r )$
(simple numbers for $k = 2$), and powers of $r$ as required 
by dimensional analysis. The suppressed terms involve subleading 
Galilei-invariant fields and noninvariant fields such as 
${\bf v}^k$, $\nabla {\bf v}^k$, etc.
By differentiating (\ref{vope}), one obtains a manifestly Galilei-invariant
OPE for velocity gradients,
\begin{eqnarray}
\nabla \cdot {\bf v}({\bf r}_1) \dots \nabla \cdot {\bf v}({\bf r}_k) 
  & \sim & \prod_{i = 2}^{k} \delta({\bf r}_i - {\bf r}_{i - 1}) 
  \Sigma_k ({\bf r}_1) + 
  \nonumber \\
  & & s_k ({\bf r}_1) + \dots \;;
\end{eqnarray}  
see also~\cite{Eyink} for general turbulent systems. Here  
the shock fields generate only contact 
singularities, while the slope fields have regular coefficients. It is
not known whether there are other singular terms.

The OPE is a consistency condition 
between renormalized correlation functions in the stationary state.
It relates, for example, the $n$-point function
$\langle {\bf v}({\bf r}_1) \dots {\bf v}({\bf r}_n) \rangle$
to the $(n - k + 1)$-point functions
$\langle \Sigma_k({\bf r}_1) {\bf v}({\bf r}_{k + 1}) 
         \dots {\bf v}({\bf r}_n)  \rangle$ and 
$\langle s_k({\bf r}_1) {\bf v}({\bf r}_{k + 1}) 
         \dots {\bf v}({\bf r}_n)  \rangle$	 
in the limit 
$ a \ll |{\bf r}_ {ij}| \ll |{\bf r}_ {il}|, r_0$ 
($i,j = 1, \dots, k$ and $l =  k+1, \dots, n$). 
The coefficient functions $C_k$ etc.~are assumed to describe local properties
of the inertial regime independent of the cutoffs $a$ and $r_0$. 
Hence, the existence of an OPE embodies two 
important characteristics of the turbulent field theory: 
(i) It is renormalized, i.e., the short-distance singularities of all
scaling fields have been removed in a consistent way. 
(ii) {\em The large-distance singularities are created solely by the
single-point amplitudes of negative-dimensional fields},
$\langle {\cal O} \rangle \sim r_0^{-x_{\cal O}}$.
In particular, the intermittency exponent $\tilde x_k$ in Eq.~(\ref{vk})
equals the scaling dimension of the leading Galilei-invariant field
in the expansion~(\ref{vope}). This is indeed the case for Burgers
turbulence, where 
$\langle u^k \rangle \sim \langle \Sigma_k \rangle \sim r_0^{- \tilde x_k}$
with $\tilde x_k = - (k - 3)/3$ according to Eqs.~(\ref{uk}) and 
(\ref{sigmak}).

The OPE (\ref{vope}) has important consequences for Burgers dynamics,
which we now discuss  for simplicity in $d = 1$.  
For purely convective dynamics ($\nu = 0$, $\varepsilon = 0$),
the moments of the excess velocity $w^k({\bf r},t)$ are locally conserved,
$\partial_t w^k + (k/(k + 1)) \nabla (w^{k + 1}) = 0$ ($k = 1,2, \dots$). 
In the driven state,
$\langle w^k ({\bf r},t) \rangle$ is pumped with a finite rate 
$\varepsilon \langle \partial_w^2 w^k ({\bf r},t) \rangle 
 \sim r_0^{- x_{k-2}}$ ($k = 2,4, \dots$). 
This cannot be offset by convection since 
$\langle \nabla (w^{k + 1})({\bf r},t) \rangle = 0$.
Hence, the stationary state must be maintained by a dynamical
anomaly~\cite{Polyakov},
\begin{equation}
\varepsilon \langle \partial_w^2 w^k ({\bf r}) \rangle 
 + \nu \langle (\nabla^2 w) ({\bf r}) \, 
               \partial_w w^k ({\bf r}) \rangle = 0 \;.
\end{equation}
It is easy to check that the OPE (\ref{vope}) predicts 
the correct form of the anomaly. We have
\begin{eqnarray}
\nu \langle (\nabla^2 w) ({\bf r}) \, w^{k-1} ({\bf r}) \rangle  
  & \sim & \nu \delta(0) \, \langle s_k ({\bf r}) \rangle 
  \nonumber \\ 
  & \sim & r_0^{-\tilde x_k + 1/3} \sim r_0^{- x_{k-2}} \;,
\end{eqnarray}
using $\nu \sim a_0^{4/3} \sim a r_0^{1/3}$ and a regularization on the
scale of the short-distance cutoff, $\delta(0) \sim a^{-1}$.
Hence, the OPE and the resulting anomalies provide 
a link between the fields $v^k$ and the shock fields $\Sigma_k$. This fixes the
intermittency exponents by the scaling relations 
$\tilde x_k = x_{k-2} + 1/3$ in accordance with (\ref{dvk}) and 
(\ref{sigmak}). 

The role of dissipation is more subtle for the 
velocity differences.
Consider again the shock number expansion 
$P_r (u) = (1-\rho) P_r^0 (u) + \rho P_r^1 (u) + O(\rho^2)$ 
for $\rho \ll 1$. By virtue of the 
OPE (\ref{vope}), the equation of motion for the 
conditional distributions $P_r^0 (u)$ and $P_r^1 (u)$ is then 
essentially reduced to that of the families of single-point amplitudes 
$\langle \Sigma_k \rangle$ and
$\langle s_k \rangle_{\rm reg}$, respectively.

{\it 1.}\ The {\em zero-shock part} $P^0_r (u)$ 
is also covariant under the `convective' scale transformations
$r \to b r$, $u \to b u$ at fixed $\varepsilon$ and $r_0$
and  can 
hence be written in the scaling form $P_r^0 (u) = (1/ u_r) {\cal P}^0 (u/u_r)$
discussed in~\cite{Polyakov}. In other words, the  expansion
$u^k = r^k s_k + (1/2) r^{k+1} \nabla s_k + \dots$
is dominated by the first term, which 
 depends on the 
large-distance scale $r_0$ only through the  ramp slope moment
$ \langle s_k \rangle \sim (u_0/r_0)^k$.
The convective symmetry is expected to be broken for 
$- u \, \raisebox{-.6ex}{$\stackrel{>}{\displaystyle{\sim}}$} 
 \, \tilde u_r \equiv (\epsilon r)^{1/3}$ by curvature effects
at preshocks. This is precisely the scale where $P_r(u)$ 
becomes dominated by the shock part; see (\ref{P01}) below. 
In the equation of motion for ${\cal P}^0$, dissipation can be neglected
at all points of finite $\nabla^2 v$ (and even at preshocks).
Furthermore, driving and convection are local processes in velocity space. It is 
straightforward to show that they are represented by the differential
operators 
$G_2 \partial_\omega^2 {\cal P}^0 (\omega)$ and 
$(\partial_\omega \omega^2 + \omega) {\cal P}^0 (\omega)$, 
respectively~\cite{Polyakov,GotohKraichnan}.
In particular, the term $\omega {\cal P}^0 (\omega)$ describes a change in
measure due to convective squeezing or stretching of the
ramps~\cite{GotohKraichnan}. We will be interested in solutions with 
a positive average ramp slope, i.e., with a net gain in measure, 
$\langle \omega \rangle > 0$. The formation of shocks, on the other 
hand, produces a measure loss $L(\omega) < 0$. In the stationary 
state, the net loss offsets the convective gain, 
$\langle \omega \rangle + \langle L(\omega) \rangle = 0$.
For consistency with known properties of $P_r (u)$, the
equation of motion 
$(G_2 \partial_\omega^2 + \partial_\omega \omega^2 + \omega 
       + L) {\cal P}^0 = 0$
must have a normalized, positive solution which behaves 
asymptotically as  
${\cal P}^0 \sim \omega \exp (- \omega^3 / 3 G_2)$ 
for $\omega \gg 1$~\cite{GurarieMigdal96,EAl}  
and
${\cal P}^0 \sim | \omega |^{-3 - \alpha}$ for $\omega \ll -1$,
with $\alpha \geq 0$, most likely $\alpha = 1/2$~\cite{EAl}. 
This requires
$L(\omega) = o(\omega)$ for $\omega \gg 1$ and
$L(\omega) \simeq \alpha \omega$  
for $\omega \ll -1$.
The `anomaly' $L(\omega)$ can be associated with ultraviolet-finite
operator products $\Sigma_1(r) s_k(r)$~\cite{Sigmas}. It is not clear,
however, whether the functions $L(\omega)$ and, hence, ${\cal P}^0 (\omega)$
are entirely determined by the OPE.

{\it 2.}\ The {\em single-shock part} $P^1_r(u)$ is expected to have 
the scaling form 
$P_r^1(u) = (\rho / u_0) {\cal P}^1 (- u / u_0) + O(\rho^2)$.
The leading term depends on $r$ only through the 
shock probability $p_1 \simeq \rho$ in accordance with the OPE
 (\ref{vope}) and the moments (\ref{uk}). 
${\cal P}^1 (\sigma)$ is the scaled shock size distribution function.
The equation of motion for a single shock produces a driving 
term $\partial_\sigma^2 {\cal P}^1 (\sigma)$ 
and a convection term 
$ \partial_\sigma f(\sigma) \sigma {\cal P}^1 (\sigma)$,
while dissipation can again be neglected. 
Here $(u_0/r_0) f(\sigma)$ is the expectation value of the average ramp
slope to both sides of a shock of size $u_0 \sigma$. 
For very large shocks ($\sigma \gg 1$), this will be positive and 
proportional to the shock size, leading to the asymptotic 
equation of motion~\cite{interactions} 
$ (\partial_\sigma^2 + A \partial_\sigma \sigma^2  ) 
  {\cal P}^1 = 0$ 
with a constant $A > 0$. 
This determines the tail of the shock size distribution,
${\cal P}^1 \sim \exp ( - A \sigma^3 / 3 )$ for $\sigma \gg 1$,
in agreement with instanton calculations~\cite{BalkovskyAl}, while 
the dynamics of initial shock growth suggests ${\cal P}^1 \sim \sigma$
for $\sigma \ll 1$~\cite{Bec}.  

The distribution $P_r(u)$ should be dominated by the zero-shock part
for small velocity differences $ |u| \sim u_r$ and by the single-shock part
for large negative values $u \sim - u_0$. 
The crossover scale 
$-\tilde u_r = -u_r \rho^{-2/3}$
is obtained by matching the two expressions, 
$P_r^0 (-\tilde u_r) \sim \rho P_r^1 (-\tilde u_r)$.
Thus we conjecture, using the
scaling form of Eq.~(\ref{calP}), 
\begin{equation}
{\cal P} (\omega, \rho) \simeq \left \{
\begin{array}{ll}
{\cal P}^0 (\omega)             & \;\; 
(\omega \rho^{2/3} \raisebox{-.6ex}{$\stackrel{>}{\displaystyle{\sim}}$} 
 - 1) 
\\
\rho^2 {\cal P}^1 (-\omega \rho) & \;\; 
(\omega \rho^{2/3} 
\raisebox{-.6ex}{$\stackrel{<}{\displaystyle{\sim}}$} 
- 1)
\end{array}                    \right . \;.
\label{P01}
\end{equation}
It is easy to verify that this solution 
%the solution given by (\ref{P01}), (\ref{P0as}), and (\ref{P1as}) 
is indeed normalizable and
has the correct moments (\ref{uk}). It has $\langle u \rangle > 0$ over
ramps and $\langle u \rangle < 0$ at shocks, which is compatible with the 
constraint $\langle u \rangle = 0$ due to translational invariance. 

To summarize: Burgers field theory contains two different families 
of local scaling fields, $\Sigma_k$ and $s_k$, which represent
powers of singular and regular velocity gradients, respectively.
The fields $\Sigma_k$ live on the shocks; their amplitudes 
$\langle \Sigma_k \rangle$ generate intermittency.
They are coupled to the other scaling fields through an OPE.
The resulting dynamical anomalies fix the intermittency exponents 
through scaling relations.
The anomalies for velocity differences have a simple 
physical cause: the formation of singular velocity configurations
out of regular ones. 

How much of this framework is preserved in Navier-Stokes turbulence can only
be conjectured at present. Intermittency is still created 
by the infrared-divergent one-point
amplitudes of negative-dimensional Galilei-invariant fields. 
These are associated with the singularities of the turbulent flow.  
Vortex filaments, for example, could play the role of the Burgers 
shocks~\cite{Frisch.book}. The singularities have a much more complicated 
statistics, however. They lead to multifractal instead of bifractal scaling
and may indeed suppress a coherent convective scaling regime 
$P_r (u) \simeq (1/u_r) {\cal P}^0 (\omega)$. 
A stronger breakdown of convective symmetry may produce anomalies and,
hence, scaling relations compatible with 
intermittency exponents $\tilde x_k$ nonlinear in $k$. Will such 
relations actually determine the values of
$\tilde x_k$, leading to a nonperturbative theory of turbulence? 

I thank Victor Yakhot for useful discussions.


\begin{references}
\bibitem{Sigmas} 
The leading term $L(\omega) \simeq \alpha \omega$ for $\omega \ll -1$
is given by the operator product $\Sigma_1(r) s_k(r) \sim s_{k+1} (r)$.
The subleading terms, however, involve the scale $r_0$ explicitly.
Polyakov's OPE~\cite{Polyakov,Boldyrev} contains only the families 
of ultraviolet-finite fields $v^k$ and $\nabla (v^k)$, or
equivalently, the generating fields 
$\exp (qv)$ and $\nabla \exp (qv)$. In terms of these fields, the 
anomaly $L(\omega)$ cannot be represented by local operator products.
\bibitem{interactions}
Both approximations neglect shock-shock interactions, which are important
at intermediate values of $\sigma$.
\end{references}
\end{document}